\RequirePackage{fix-cm}
\documentclass[twocolumn,epjc3]{svjour3}  
\smartqed  
\RequirePackage{graphicx}
\RequirePackage{bm}
\RequirePackage{mathrsfs}
\RequirePackage{amsmath}
\RequirePackage{cancel}
\RequirePackage{amssymb}
\RequirePackage{url}
\RequirePackage{amsmath}
\RequirePackage{cases}
\begin{document}

\title{Gravitomagnetism and Gravitational Waves in Galileo-Newtonian Physics}



\author{Harihar Behera\thanksref{e1}}
%

\thankstext{e1}{e-mail: behera.hh@gmail.com}


\institute{Physics Department, BIET Higher Secondary School, Govindpur, Dhenkanal-759001, Odisha, India}

\date{Received: date / Accepted: date}
\maketitle
\begin{abstract}
Adopting Schwinger's formalism for inferring Maxwell-Lorentz equations (MLEs) and combining three ingredients: (i) the laws of gravitostatics, (ii) the Galileo-Newton principle of relativity 
and (iii) existence of gravitational waves which travel in vacuum with a finite speed $c_g$, we inferred two sets of gravito-Maxwell-Lorentz Equations (g-MLEs). One of these sets corresponds to Heaviside's Gravity of 1893 and the other set corresponds to what we call Maxwellian Gravity (MG). HG and MG are mere two mathematical representations of a single physical theory called Heaviside-Maxwellian Gravity (HMG). While rediscovering Heaviside's gravitational field equations following Schwinger's formalism, we found a correction to Heaviside's speculative gravito-Lorentz force law. This work presents a Galilo-Newtonian foundation of gravitomagnetic effects and gravitational waves, caused by time-varying sources and fields, which are currently considered outside the domain of Newtonian physics.  The emergence HMG from other well-established principles of physics is also noted, which established its theoretical consistency and fixed the value of $c_g$ uniquely at the speed of light in vacuum. The explanations of certain experimentally verified general relativistic results within HMG are also noted. We also report, a set of new Maxwell-Lorentz Equations, physically equivalent to the standard set, as a byproduct product of the present approach.

\keywords{Gravitomagnetism \and Gravitational Waves \and Speed of Gravitational Waves \and Heaviside's Vector Gravity} 
\PACS{03.50.kk \and 04.20.Cv \and 04.30.Db \and 04.30.Nk }
\end{abstract}
\section{Introduction}
\label{Intro.}
In a remarkable formalism, Schwinger et al. \cite{1} have obtained the Maxwell-Lorentz equations (MLEs) of electrodynamics by combining three ingredients: the laws of electrostatics, the Galileo-Newton principle of relativity (charges at rest, and charges with a common velocity viewed from a co-moving observer, are physically indistinguishable); and the existence of electromagnetic waves that travel in a vacuum at the some finite speed $c$. In the discussion of their inference of MLEs, Schwinger et al. \cite{1} stated, ``\emph{Einsteinian relativity is an outgrowth of of Maxwellian electrodynamics, not the other way about. That is the spirit in which electrodynamics is developed as a self-consistent subject in this book.}" \\
Motivated by the striking formal analogy between the Newton's laws of gravitostatics and the Coulomb's laws of electrostatics and essentially adopting Schwinger's formalism, here we show how one can extend the gravito-static equations of Newtonian gravity to time dependent sources and fields to obtain the basic time-dependent equations of gravitodynamics of moving bodies within the Galileo-Newtonian physics by combining the following three plausible assumptions:  \\
(A) Newton's laws of gravitostatics are valid for time-independent sources and fields, \\
(B) The local conservation of mass expressied by the equation of continuity is a 
valid law\footnote{Following Schwinger et al.\cite{1}, the assumption (B) may be derived from the Galilean   principle of relativity (the rest mass is an invariant quantity under the transformation laws of Galilio-Newtonian Relativity (GNR)). However, the relativity principle of GNR will be used here for  obtaining the analogue of the Lorentz force law for gravitation.},  and \\
(C) Time-dependent sources and fields in Newtonian gravity produce gravitational waves that travel through  vacuum with a universal finite velocity $c_g$, the value of which may be obtained from measurement of the physical quantities involving $c_g$ or from a comparison of its field equations with those obtainable from more advanced theories like special relativistic gravity or general relativistic gravity reduced to flat space time, i.e. linearized equations of Einstein's gravity). \\
As we shall see here, we will obtain two physically equivalent sets of gravito-Maxwell-Lorentz equations as listed in Table 1. One of the two sets represents what we call Maxwellian Gravity (MG)\cite{2,3,4} and the other set represents Heaviside Gravity (HG) \cite{5,6,7,8,9,10,11} in its original form\footnote{Heaviside's original speculative gravito-Lorentz force was: $\frac{d\mathbf{p}}{dt}\,=\,m_g\left[\mathbf{g}\,+\,\mathbf{u}\times \mathbf{b}\right]$ \cite{6,7,8,9,10,11}; but corrected here as in Table 1: $\frac{d\mathbf{p}}{dt}\,=\,m_g\left[\mathbf{g}\,-\,\mathbf{u}\times \mathbf{b}\right]$.} (represented here in our notation and convention). 
\begin{table}[h]
\begin{center}
   \begin{tabular}{ | l | l |}
    \hline
    gravito-MLEs of MG & gravito MLEs of HG \\ \hline
    $\mathbf{\nabla}\cdot\mathbf{g} = - 4\pi G \rho_g = - \frac{\rho_g}{\epsilon_{0g}}$ & $\mathbf{\nabla}\cdot\mathbf{g} = - 4\pi G \rho_g= - \frac{\rho_g}{\epsilon_{0g}}$ \\ \hline
    $\mathbf{\nabla}\cdot\mathbf{b} = 0$ & $\mathbf{\nabla}\cdot\mathbf{b} = 0$ \\ \hline
    $\mathbf{\nabla}\times\mathbf{b}\,=\,-\,\mu_{0g}\mathbf{j}_g\,+\,\frac{1}{c_g^2}\frac{\partial \mathbf{g}}{\partial t}$ & $\mathbf{\nabla}\times\mathbf{b}\,=\,+\,\mu_{0g}\mathbf{j}_g\,-\,\frac{1}{c_g^2}\frac{\partial \mathbf{g}}{\partial t}$  \\ \hline 
    $\mathbf{\nabla}\times\mathbf{g}\,=\,-\,\frac{\partial \mathbf{b}}{\partial t}$ & $\mathbf{\nabla}\times\mathbf{g}\,=\,+\,\frac{\partial \mathbf{b}}{\partial t}$  \\ \hline
    $\frac{d\mathbf{p}}{dt}\,=\,m_g\left[\mathbf{g}\,+\,\mathbf{u}\times \mathbf{b}\right]$ & $\frac{d\mathbf{p}}{dt}\,=\,m_g\left[\mathbf{g}\,-\,\mathbf{u}\times \mathbf{b}\right]$  \\ \hline 
   $\mathbf{b}\,=\,+\mathbf{\nabla}\times \mathbf{A}_g$ & $\mathbf{b}\,=\,-\mathbf{\nabla}\times \mathbf{A}_g$ \\ \hline
     $\mathbf{g}\,=\,-\,\mathbf{\nabla}\phi_{g} \,-\,\frac{\partial \mathbf{A}_g}{\partial t}$ & $\mathbf{g}\,=\,-\,\mathbf{\nabla}\phi_{g} \,-\,\frac{\partial \mathbf{A}_g}{\partial t}$ \\ \hline
\end{tabular}
\end{center}
\caption{Physically Equivalent Sets of gravito-Maxwell-Lorentz Equations (g-MLEs) representing Heaviside-Maxwellian Gravity.}
    \label{tab:truthTables}   
\end{table}
In Table 1, in analogy with Maxwell-Lorentz equations of classical electromagnetism in SI units, we have introduced two new universal constants for vacuum: 
\begin{equation}
\label{eq:1}
\epsilon_{0g} = \frac{1}{4\pi G}, \quad \mu_{0g} = \frac{4\pi G}{c_g^2} \quad \Rightarrow \quad c_g = \frac{1}{\sqrt{\epsilon_{0g}\mu_{0g}}}
\end{equation}
where $c_g$ is the universal finite propagation velocity of gravitational waves in vacuum, $G$ is the Newton's gravitational constant, $\rho_g = \rho_0$ is the positive ordinary (rest) mass density (which is also shown to be valid in relativistic case in ref. \cite{4} without sacrificing the time-tested empirical law of universality of free fall of Galileo), $\mathbf{j}_g = \rho_0\mathbf{v}$ stands for the mass current density ($\mathbf{v}$ is velocity) and by electromagnetic analogy, $\mathbf{b}$ may be called the gravito-magnetic field, the Newtonian gravitational field $\mathbf{g}$ may be called as gravito-electric  field, $\epsilon_{0g}$ and $\mu_{0g}$ may be named respectively as the gravito-electric (or gravitic) permitivity and the gravito-magnetic permeability of vacuum. Because of their physical equivalence, both HG and MG represent the same physical theory that was named recently as Heaviside-Maxwellan Gravity by Behera and Barik \cite{12}, who obtained the fundamental equations of MG and HG as listed in Table 1 (with $c_g = c,$ the propagation speed of light in vacuum) using the principle of local phase (or gauge) invariance of quantum field theory. 

The present Galileo-Newtonian approach has relevance and significance in the present era of experimental detection of gravitational waves \cite{13,14,15,16}, gravito-magnetic field of the spinning Earth by (i) analising the orbital dynamics and parameters of two laser-ranged satellites (LAGEOS and LAGEOS II) \cite{17,18,19,20,21} and (ii) by measuring, very precisely, tiny changes in the directions of the spin axis of four gyroscopes contained in a polar Earth satellite in The Gravity Probe B (or GP-B) Experiment
 \cite{22,23,24}, which is  a  NASA  physics  mission planned to experimentally investigate Einstein's 1916 theory of gravity. These new experimental results are being interpreted in the literature as new crucial tests confirming the predictions of general relativity, which have no Galileo-Newtonian counterparts.
For instance, Ciufolini et al. \cite{17}, who is leading the orbital data analysis of LAGEOS  (LAser GEOdynamics Satellite) and LAGEOS II satellites, stated,
\begin{quote}
Newton's law of gravitation has a formal counterpart in Coulomb's law of electrostatics; however, Newton's theory has no phenomenon formally analogous to magnetism. On the other hand, Einstein's theory of gravitation predicts that the force generated by a current of electrical charge, described by Amp\'{e}re's law, should also have a formal counterpart ``force" generated by a current of mass. The detection and measurement of this ``gravitomagnetic" force is the subject of this report.
\end{quote}
\section{Inference of gravito-Maxwell-Lorentz Equations (g-MLEs)}
According to the Newton's laws of gravitostatics, the source of gravitostatic or gravitoelectric field $\mathbf{g}$ is the gravitational mass $m_g = m_0$ (or proper/rest mass\footnote{The equality $m_g = m_0$ is true in Galileo-Newtonian physics for the motion of a body in a gravitational field to be independent of its rest mass. This equality is also valid in relativistic physics as demonstrated in \cite{4} using the principle of Lorentz in-variance of physical laws.}). 
The gravito-electric field $\mathbf{g}$, in gravitostatics, satisfies the following two equations, viz., 
\begin{equation}\label{eq:2}
\mathbf{\nabla}\cdot\mathbf{g}\,=\,-\,4\pi G\rho_g = - \rho_0/\epsilon_{0g} \qquad {\mbox{and}}
\end{equation}
\begin{equation}\label{eq:3}
\mathbf{\nabla}\times \mathbf{g}\,=\,\mathbf{0}, 
\end{equation}
\noindent 
The gravitostatic force on a point mass $m_0$ in a gravito-static field $\mathbf{g}$ is expressed by the following law (valid in an inertial frame)  
\begin{equation}\label{eq:4}
\mathbf{F}_g = m_0\mathbf{g} = - m_0 \mathbf{\nabla}\phi_g,  
\end{equation}
\noindent
where $\phi_g$ is the scalar potential at the position of $m_0$. 
The local conservation law for (proper) mass is expressed by the continuity equation:
\begin{equation}\label{eq:5}
\mathbf{\nabla}\cdot \mathbf{j}(\mathbf{r}, t)\,+\,\frac{\partial }{\partial t}\rho_0(\mathbf{r},t)\,=\,0,
\end{equation} 
\noindent
where $\mathbf{j} = \rho_0(\mathbf{r},t) \mathbf{v}$ represents the mass current density. In Galileo-Newtonian physics, the individual validity of the three laws represented by eqs. \eqref{eq:2}, \eqref{eq:3} and \eqref{eq:5} is indisputable. But these eqs. \eqref{eq:2}, \eqref{eq:3} and \eqref{eq:5} have their simultaneous validity only for the systems or situations where the following three equations, viz.,
\begin{subequations}\label{eq:6}
\begin{equation}
\label{eq:6:1}
\mathbf{\nabla}\cdot \mathbf{j}(\mathbf{r}, t)\,=\,0,
\end{equation}
\begin{equation}
\label{eq:6:2}
\frac{\partial }{\partial t}\rho_0(\mathbf{r},t)\,=\,0,
\end{equation} 
\begin{equation}
\label{eq:6:3}
\frac{\partial }{\partial t}\mathbf{g}(\mathbf{r}, t)\, =\,\mathbf{0}.
\end{equation}  
\end{subequations}
 \noindent
remain valid\footnote{In eqs. \eqref{eq:6}, we have made the time dependence of $\mathbf{j}(\mathbf{r}, t)$, $\rho_0(\mathbf{r},t)$ and $\mathbf{g}(\mathbf{r},t)$ explicit.}. The eqs. \eqref{eq:2}-\eqref{eq:6} taken with Newton's second law of motion describe the dynamics of masses in gravitostatics. \\
Newton's gravitational law \eqref{eq:4} suggests is action-at-a-distance: A mass at one point acts directly and instantaneously on another mass even when the two masses are not it contact but widely separated in space like the Sun and the Earth. Although Newton was aware of this problem, he could not resolved it \cite{4}. Here, we wish to resolve, the above mentioned Newwton's unresolved problem, within his domain of physics without any appeal to Einstein's relativity. To this aim, we first consider a  system where the Gauss's law of gravitostatics \eqref{eq:2} and the equation of continuity \eqref{eq:5}  
co-work simultaneously, but now with the restrictions in eq. \eqref{eq:6} removed by imposing three conditions: viz., 
\begin{subequations} \label{eq:7}
\begin{align}
\mathbf{\nabla}\cdot \mathbf{j}(\mathbf{r}, t)\,\neq \,0, \\
\frac{\partial }{\partial t}\rho_0(\mathbf{r},t)\,\neq \,0, \\
\frac{\partial }{\partial t}\mathbf{g}(\mathbf{r}, t)\, \neq \, \mathbf{0}.
\end{align}
\end{subequations}
With these conditions, now we take the time derivative of \eqref{eq:2} and then write the result as
\begin{equation}\label{eq:8}
\frac{\partial \rho_0}{\partial t}\,=\,-\,\epsilon_{0g} \mathbf{\nabla}\cdot \frac{\partial \mathbf{g}}{\partial t}.
\end{equation}
Now using eq. \eqref{eq:8} in eq. \eqref{eq:5} we obtain 
\begin{equation}\label{eq:9}
\mathbf{\nabla}\cdot\left(\mathbf{j}\,-\,\epsilon_{0g} \frac{\partial \mathbf{g}}{\partial t}\right)\,=\,0.
\end{equation}
The quantity inside the parenthesis of \eqref{eq:9} is a vector whose divergence is zero. Since $\mathbf{\nabla}\cdot(\mathbf{\nabla}\times\mathbf{A})\,=\,0$\, for any vector $\mathbf{A}$, we express 
the vector inside the parenthesis of \eqref{eq:9} as the curl of some vector, say $\mathbf{h}$. Now we find that Eq. \eqref{eq:9} is mathematically satisfied by two independent representations of $\mathbf{h}$:
\begin{align} \label{eq:10}
\mathbf{\nabla}\times\mathbf{h}\,&= \,\pm \left(\mathbf{j}\,-\,\epsilon_{0g}\frac{\partial \mathbf{g}}{\partial t}\right) \nonumber \\
         &= \begin{cases} \,+\,\mathbf{j}\,-\,\epsilon_{0g}\frac{\partial \mathbf{g}}{\partial t} & \quad \text{(For HG)} \\ 
\,-\,\mathbf{j}\,+\,\epsilon_{0g}\frac{\partial \mathbf{g}}{\partial t} & \quad \text{(For MG)}
\end{cases}
\end{align}
Now we multiply a new non-zero finite and positive valued dimensional scalar constant $\mu_{0g}$ with eq. \eqref{eq:10} to get the following equation \footnote{Such multiplication does not alter the physical content of a vector equation, since it only re-scales physical quantities involved and re-express them in some new units.}
\begin{equation} \label{eq:11}
\mathbf{\nabla}\times\mathbf{b}\,=  \begin{cases} \,+\,\mu_{0g} \mathbf{j}\,-\,\epsilon_{0g} \mu_{0g}\frac{\partial \mathbf{g}}{\partial t} & \quad \text{(For HG)} \\ 
\,-\,\mu_{0g}\mathbf{j}\,+\,\epsilon_{0g} \mu_{0g}\frac{\partial \mathbf{g}}{\partial t} & \quad \text{(For MG)}
\end{cases}
\end{equation}
such that the $ \sqrt{\epsilon_{0g} \mu_{0g}} = c_g^{-1}$; $c_g$, having the dimensions of speed, would  emerge as the speed of gravitational waves in free space and $\mathbf{b}= \mu_{0g}\mathbf{h}$ for free space. In vacuum, where $\mathbf{j} = \mathbf{0}$, eqs. \eqref{eq:11} become 
\begin{equation} \label{eq:12}
\mathbf{\nabla}\times\mathbf{b}\,=  \begin{cases} \,-\,\frac{1}{c_g^2} \frac{\partial \mathbf{g}}{\partial t} & \quad \text{(For HG)} \\ 
\,+\,\frac{1}{c_g^2}\frac{\partial \mathbf{g}}{\partial t} & \quad \text{(For MG)}
\end{cases}
\end{equation}
Taking the curl of the eqs. \eqref{eq:12}, we get 
\begin{align} \label{eq:13}
\mathbf{\nabla}\times(\mathbf{\nabla}\times\mathbf{b})&= \mathbf{\nabla}(\mathbf{\nabla}\cdot \mathbf{b}) - \mathbf{\nabla}^2 \mathbf{b}  \nonumber \\
         &= \begin{cases} - \frac{1}{c_g^2} \frac{\partial }{\partial t} \mathbf{\nabla}\times \mathbf{g} & \quad \text{(For HG)} \\ 
+ \frac{1}{c_g^2} \frac{\partial }{\partial t} \mathbf{\nabla}\times \mathbf{g} & \quad \text{(For MG)}
\end{cases}
\end{align}  
Equations \eqref{eq:13}, reduce to the wave equations for the $\mathbf{b}$ field, viz., 
\begin{equation}\label{eq:14}
\mathbf{\nabla}^2\mathbf{b}\,=\,\frac{1}{c_g^2}\frac{\partial ^2 \mathbf{b}}{\partial t^2} \quad \text{(For both HG and MG)}
\end{equation}
\noindent
under the following two conditions:
\begin{equation}\label{eq:15}
\mathbf{\nabla}\cdot \mathbf{b} = 0 \quad \text{(For both HG and MG)}
\end{equation}
\begin{align}\label{eq:16}
\mathbf{\nabla}\times\mathbf{g} &= \begin{cases} + \frac{\partial \mathbf{b}}{\partial t}  & \quad \text{(For HG)} \\ 
- \frac{\partial \mathbf{b}}{\partial t} & \quad \text{(For MG)}
 \end{cases}
\end{align}
Now taking the curl of eqs. \eqref{eq:16}, we get 
\begin{align} \label{eq:17}
\mathbf{\nabla}\times(\mathbf{\nabla}\times\mathbf{g})&= \mathbf{\nabla}(\mathbf{\nabla}\cdot \mathbf{g}) - \mathbf{\nabla}^2 \mathbf{g}  \nonumber \\
         &= \begin{cases} + \frac{\partial }{\partial t} \mathbf{\nabla}\times \mathbf{b} & \quad \text{(For HG)} \\ 
- \frac{\partial }{\partial t} \mathbf{\nabla}\times \mathbf{b} & \quad \text{(For MG)}
\end{cases}
\end{align}
In vacuum $\mathbf{\nabla}\cdot \mathbf{g} = 0$ and $\mathbf{\nabla}\times \mathbf{b}$ is given by eqs. \eqref{eq:12}, so eqs. \eqref{eq:17} give us the wave equation for the $\mathbf{g}$ field, viz., 
\begin{equation}\label{eq:18}
\mathbf{\nabla}^2\mathbf{g}\,=\,\frac{1}{c_g^2}\frac{\partial ^2 \mathbf{g}}{\partial t^2} \quad \text{(For both HG and MG)}.
\end{equation}
Since $\mathbf{\nabla}\cdot \mathbf{b} = 0 $ for both HG and MG as seen in eq. \eqref{eq:15} and 
$\mathbf{\nabla}\cdot(\mathbf{\nabla}\times\mathbf{A})\,=\,0$\, for any vector $\mathbf{A}$, the $\mathbf{b}$ field can be expressed as the curl of some vector function $\mathbf{A}_g$  (say). If we adopt the following definitions, viz., 
\begin{align}\label{eq:19}
\mathbf{b} &= \begin{cases} - \mathbf{\nabla}\times\mathbf{A}_g  & \quad \text{(For HG)} 
\\ 
+ \mathbf{\nabla}\times\mathbf{A}_g & \quad \text{(For MG)},
 \end{cases}
\end{align}
with $\mathbf{A}_g$ being the vector potential, then in view of the gravito-Faraday law expressed in eq. \eqref{eq:16}, the $\mathbf{g}$ field can be expressed in terms of $\phi_g$ and $\mathbf{A}_g$ as 
\begin{equation}\label{eq:20}
\mathbf{g}\,=\,-\,\mathbf{\nabla}\phi_g\,-\,\frac{\partial \mathbf{A}_g}{\partial t} \qquad \text{(For both HG and MG)}.
\end{equation}
Now substituting eq. \eqref{eq:20} and \eqref{eq:19} in eqs. \eqref{eq:11}, we get the following expressions for the in-homogeneous eqs. \eqref{eq:11} in terms of potentials $(\phi_g, \mathbf{A}_g)$ as
\begin{equation} \label{eq:21}
\mathbf{\nabla}^2\phi_g - \frac{1}{c_g^2} \frac{\partial ^2\phi_g}{\partial t^2}=
\frac{\rho_0}{\epsilon_{0g}} \quad \text{(For both HG \& MG)}, 
\end{equation}
\begin{equation} \label{eq:22}
\mathbf{\nabla}^2\mathbf{A}_g - \frac{1}{c_g^2} \frac{\partial ^2\mathbf{A}_g}{\partial t^2}
=\mu_{0g}\mathbf{j} \quad \text{(For both HG \& MG)}, 
\end{equation}
\noindent
by imposing the following gravito-Lorenz gauge condition, 
\begin{equation} \label{eq:23}
\mathbf{\nabla}\cdot\mathbf{A}_g\,+\,\frac{1}{c_g^2}\frac{\partial \phi_g}{\partial t}\,=\,0 \qquad \text{(For both HG and MG)}.
\end{equation}
As in classical electrodynamics, the generation of gravitational waves by prescribed mass and mass current distributions will be determined by the eqs. \eqref{eq:21}-\eqref{eq:22}. The retarded solutions of eq. \eqref{eq:21} and \eqref{eq:22} are, respectively, 
\begin{equation}\label{eq:24}
\phi_g(\mathbf{r}, t) = \,-\,\frac{1}{4\pi \epsilon_{0g}}\int \frac{\rho_g(\mathbf{r}^\prime , t^\prime)}{|\mathbf{r} -\mathbf{r}^\prime|}dv^\prime \quad {\mbox{and}}
\end{equation}
\begin{equation}\label{eq:25}
\mathbf{A}_g(\mathbf{r}, t) = \,-\,\frac{\mu_{0g}}{4\pi }\int \frac{\mathbf{j}(\mathbf{r}^\prime , t^\prime)}{|\mathbf{r} -\mathbf{r}^\prime|}dv^\prime,
\end{equation}
\noindent
where $t^\prime = t - |\mathbf{r} - \mathbf{r}^\prime|/c_g$ is the retarded time and $dv^\prime$ is an elementary volume element at $\mathbf{r}^\prime$. The reader can now realize that we have arrived at two sets of gravito-Maxwell's equations noted in Table 1, that can produce gravitational waves in the spirit of Faraday and Maxwell. \\
The electromagnetic analogy, we just uncovered, suggests a modification of the force law \eqref{eq:4} to include the gravito-magnetic interaction between moving masses analogous to the magnetic interaction between two moving charges. This will complete the dynamical picture if we could find a gravitational analogue of Lorentz force law. To this end, below we adopt the formalism of Schwinger et al. \cite{1} in their derivation of the Lorentz force law. \\

Consider two inertial frames $S$ and $S^\prime$ in relative motion. Let the relative velocity of $S$ and $S^\prime$ be $\mathbf{v}$. To introduce the time-dependence of $\rho_0$ and $\mathbf{g}$ in a simplest way, suppose that all the masses are in static arrangement in one of these frames, say $S^\prime$, which is moving with velocity $\mathbf{v}$ with respect to $S$. Thus all the masses in $S^\prime$ are moving with a common velocity $\mathbf{v}$ with respect to $S$. Here we use Galilio-Newtonian principle of relativity (in-variance of rest mass under Galilean transformation: masses at rest and masses with a common velocity viewed by a co-moving observer are physically indistinguishable) and demand that physical laws remains the same in the two inertial frames. Further, assume that $|\mathbf{v}| << c$. A co-moving observer with the moving masses has to move with velocity $\mathbf{v}$ with respect to $S$. Therefore, the time derivative in the co-moving system, in which the masses are at rest, is the sum of explicit time dependent and co-ordinate dependent contributions,
 \begin{equation} \label{eq:26}
 \frac{d}{dt}\,=\,\frac{\partial}{\partial t}\,+\,\mathbf{v}\cdot \mathbf{\nabla}
\end{equation} 
so, in going from static system to uniformly moving system, one has to make the replacement 
\begin{equation}\label{eq:27}
\frac{\partial}{\partial t}\, \longrightarrow \,\frac{d}{dt}\,=\,\frac{\partial}{\partial t}\,+\,\mathbf{v}\cdot \mathbf{\nabla}.
\end{equation} 
In the moving system, the gravito-static field in equation \eqref{eq:6:3} becomes 
\begin{equation} \label{eq:28}
\mathbf{0} = \frac{\partial \mathbf{g}}{\partial t}\,\longrightarrow \,\mathbf{0}\,=\,\frac{d\mathbf{g}}{dt}\,=\,\frac{\partial \mathbf{g}}{\partial t}\,+\,(\mathbf{v}\cdot \mathbf{\nabla})\mathbf{g}.
\end{equation}
For constant $\mathbf{v}$, we have the vector identity: 
\begin{equation}\label{eq:29}
\mathbf{\nabla}\times (\mathbf{v}\times \mathbf{g})= \mathbf{v}(\mathbf{\nabla}\cdot\mathbf{g}) - (\mathbf{v}\cdot \mathbf{\nabla})\mathbf{g}.
\end{equation}
Now using Gauss's law \eqref{eq:2} in eq. \eqref{eq:29}, we get 
\begin{equation}\label{eq:30}
\begin{split}
\mathbf{\nabla}\times (\mathbf{v}\times \mathbf{g})&= - \frac{(\rho_0 \mathbf{v})}{\epsilon_{0g}}-(\mathbf{v}\cdot  \mathbf{\nabla})\mathbf{g}  \\
                                                     &= - \frac{\mathbf{j}}{\epsilon_{0g}}-(\mathbf{v}\cdot \mathbf{\nabla})\mathbf{g}.
\end{split}
\end{equation}
Substituting the value of $(\mathbf{v}\cdot \mathbf{\nabla})\mathbf{g}$ from eq. \eqref{eq:28} in eq. \eqref{eq:30}, we get 
\begin{equation}\label{eq:31}
\mathbf{\nabla}\times (\mathbf{v}\times \mathbf{g})=\,-\,\frac{\mathbf{j}}{\epsilon_{0g}} + \frac{\partial \mathbf{g}}{\partial t}.
\end{equation}
Multiplication of eq. \eqref{eq:31} by $c_g^{-2}$ gives us 
\begin{equation}\label{eq:32}
\mathbf{\nabla}\times \left(\frac{\mathbf{v}\times \mathbf{g}}{c_g^2}\right)\,=\,-\,\mu_{0g}\mathbf{j} + \frac{1}{c_g^2}\frac{\partial \mathbf{g}}{\partial t}.
\end{equation} 
Eq. \eqref{eq:32} will agree with the eqs. \eqref{eq:11} only when 
\begin{align}\label{eq:33}
\mathbf{b} &= \begin{cases} - \frac{\mathbf{v}\times \mathbf{g}}{c_g^2}  & \quad \text{(For HG)} 
\\ 
+ \frac{\mathbf{v}\times \mathbf{g}}{c_g^2} & \quad \text{(For MG)},
 \end{cases}
\end{align}
For the $\mathbf{b}$ field, consider the vector identity for constant $\mathbf{v}$: 
\begin{equation}\label{eq:34}
\mathbf{\nabla}\times (\mathbf{v}\times \mathbf{b})= \mathbf{v}(\mathbf{\nabla}\cdot\mathbf{b})\,-\,(\mathbf{v}\cdot \mathbf{\nabla})\mathbf{b} =\,-\,(\mathbf{v}\cdot \mathbf{\nabla})\mathbf{b} 
\end{equation}
where we have used $\mathbf{\nabla}\cdot\mathbf{b}\,=\,0$. Furthermore, in the co-moving system (where the masses are at rest - static) the $\mathbf{b}$ field should also not change with time:
\begin{equation}\label{eq:35}
\frac{d\mathbf{b}}{dt}\,=\,\frac{\partial \mathbf{b}}{\partial t}\,+\,(\mathbf{v}\cdot \mathbf{\nabla})\mathbf{b}\,=\,\mathbf{0}.
\end{equation}
Eqs. \eqref{eq:34} and \eqref{eq:35} give us  
\begin{equation} \label{eq:36}
\frac{\partial \mathbf{b}}{\partial t}\,=\,\mathbf{\nabla}\times (\mathbf{v}\times \mathbf{b}).
\end{equation} 
Again, the vector identity
\begin{equation} \label{eq:37}
\mathbf{\nabla}^2 \mathbf{g}\,= \,\mathbf{g}(\mathbf{\nabla}\cdot \mathbf{g})\,-\,\mathbf{\nabla}\times (\mathbf{\nabla}\times \mathbf{g})
\end{equation}
in vacuum (i.e. outside the mass distribution where $\mathbf{\nabla}\cdot \mathbf{g}\,=\,0$) becomes
\begin{equation}\label{eq:38}
\mathbf{\nabla}^2 \mathbf{g} = - \mathbf{\nabla}\times(\mathbf{\nabla}\times 
\mathbf{g}).
\end{equation}
This gives us 
the left hand side of the wave eq. \eqref{eq:18} for $\mathbf{g}$ in vacuum.
Taking the time derivative of the eq. \eqref{eq:12} and considering eq. \eqref{eq:36}, the right side of the wave eq. \eqref{eq:18} becomes 
\begin{align}\label{eq:39}
\frac{1}{c_g^2}\frac{\partial ^2 \mathbf{g}}{\partial t^2} &= \begin{cases} - \mathbf{\nabla}\times \left[\mathbf{\nabla}\times (\mathbf{v}\times \mathbf{b})\right]  & \,\, \text{(For HG)} 
\\ 
+ \mathbf{\nabla}\times \left[\mathbf{\nabla}\times (\mathbf{v}\times \mathbf{b})\right] & \,\, \text{(For MG)}
 \end{cases}
\end{align}
Eqs. \eqref{eq:38} and \eqref{eq:39} reveal that the wave eq. \eqref{eq:18} for the $\mathbf{g}$ field will hold if 
\begin{align}\label{eq:40}
\mathbf{g} &= \begin{cases} \,+\,\mathbf{v}\times \mathbf{b} & \quad \text{(For HG)} 
\\ 
\,-\,\mathbf{v}\times \mathbf{b} & \quad \text{(For MG)}.
 \end{cases}
\end{align}
\noindent
Notice that in eq. \eqref{eq:40}, as $\mathbf{v}\rightarrow \mathbf{0} \Rightarrow \mathbf{g} \rightarrow \mathbf{0}$. No gravitostatics! So eq. \eqref{eq:40} cannot be completely right. However, all that is necessary is that $\mathbf{\nabla}\times \mathbf{g}$ in eq. \eqref{eq:40} should be valid:
\begin{align}\label{eq:41}
\mathbf{\nabla}\times \mathbf{g} &= \begin{cases} \,+\,\mathbf{\nabla}\times (\mathbf{v}\times \mathbf{b}) & \quad \text{(For HG)} 
\\ 
\,-\,\mathbf{\nabla}\times (\mathbf{v}\times \mathbf{b}) & \quad \text{(For MG)}.
 \end{cases}
\end{align}
\noindent
or, if we use eq. \eqref{eq:36},
\begin{align}\label{eq:42}
\mathbf{\nabla}\times \mathbf{g} &= \begin{cases} \,+\,\frac{\partial \mathbf{b}}{\partial t} & \quad \text{(For HG)} 
\\ 
\,-\,\frac{\partial \mathbf{b}}{\partial t} & \quad \text{(For MG)},
 \end{cases}
\end{align}
which is consistent with gravitostatics since it generalizes $\mathbf{\nabla}\times \mathbf{g}\,=\,\mathbf{0}$ to time-dependent fields. \\
\noindent
Now we have all the necessary equations that are needed to address the question: What replaces the equation $\mathbf{F}_g\,=\,m_0\mathbf{g}$ to describe the force a point mass $m_0$, when that point mass moves with some non-relativistic velocity $\mathbf{v}$ in some given $\mathbf{g}$ and $\mathbf{b}$ field? 
To find an answer to this question, let us consider two inertial coordinate systems $S$ and $S^\prime$ with relative velocity $\mathbf{v}$between them. Suppose $m_0$ is at rest in $S^\prime$ (called co-moving system) which moves at velocity $\mathbf{v}$ with respect to $S$. So the velocity of $m_0$ in 
$S$ is $\mathbf{v}$. In the $S$ coordinate system, let the gravito-electric and gravito-magnetic fields are given by $\mathbf{g}$ and $\mathbf{b}$, respectively. In the co-moving system $S^\prime$, the force on $m_0$ is  
\begin{equation}\label{eq:43}
\mathbf{F}_g\,=\,m_0\mathbf{g}^{\mbox{eff}},
\end{equation}
\noindent 
where $\mathbf{g}^{\mbox{eff}}$ is the gravito-electric field in $S^\prime$. In transforming to the co-moving frame, all the other masses - those responsible for $\mathbf{g}$ and $\mathbf{b}$ - have been given an additional counter velocity $-\mathbf{v}$. From eq.  \eqref{eq:40}, we then infer that in the co-moving frame 
\begin{enumerate}
\item[(i)] ($-\mathbf{v}\times\mathbf{b}$) has the character of an additional gravito-electric field  for the case of HG, and similarly
\item[(ii)] ($+\mathbf{v}\times\mathbf{b}$) has the character of an additional gravito-electric field for the case of MG.
\end{enumerate}
\noindent
Hence, the suggested $\mathbf{g}^{\mbox{eff}}$ is 
\begin{align}\label{eq:44}
\mathbf{g}^{\mbox{eff}}\,&= \begin{cases} \,\mathbf{g}\,-\,\mathbf{v}\times\mathbf{b}  & \quad \text{(For HG)} 
\\ 
\,\mathbf{g}\,+\,\mathbf{v}\times\mathbf{b}  & \quad \text{(For MG)}.
 \end{cases}
\end{align}
\noindent
leading to the gravito-Lorentz force law of HG and MG denominations: 
\begin{align}\label{eq:45}
\mathbf{F}_{\mbox{gL}}\,&= \begin{cases} \,m_0\left(\mathbf{g}\,-\,\mathbf{v}\times\mathbf{b}\right). & \quad \text{(For HG)} 
\\ 
\,m_0\left(\mathbf{g}\,+\,\mathbf{v}\times\mathbf{b}\right). & \quad \text{(For MG)}.
 \end{cases}
\end{align}
\noindent
In an inertial frame, this force law \eqref{eq:45} when used in Newton's 2nd law of motion, 
\begin{align}\label{eq:46}
\frac{d\mathbf{p}}{dt} = \mathbf{F}_{\mbox{gL}}&= \begin{cases} m_0\left(\mathbf{g}\,-\,\mathbf{v}\times\mathbf{b}\right) & \,\,\, \text{(For HG)} 
\\ 
m_0\left(\mathbf{g}\,+\,\mathbf{v}\times\mathbf{b}\right) & \,\,\, \text{(For MG)}.
 \end{cases}
\end{align}
we get the equation of a point mass $m_0$ moving with momentum $\mathbf{p}$ in the gravito-electric and gravito-magnetic filed of HG and MG denomination. With this, we completed our inference of two physically equivalent but mathematically different representations of gravito-Maxwell-Lorentz equations classical gravitoelectromagnetic (GEM) theory listed in Table 1. The value of $c_g$ is uniquely fixed at $c_g = c$ (the speed of light in vacuum) in ref. \cite{4} demanding the Lorentz in-variance of physical laws. Our present results on Maxwellian Gravity perfectly agree with several previous theoretical  \cite{2,4,11,12,25,26,26,27,28,29,30,31,32,33,34,35,36} as well as experimental studies   
 \cite{37,38,39,40,41,42} on MG if $c_g = c$.   
\section{An Set of New Maxwell-Lorentz Equations (N-MLEs) from Schwinger's Formalism} 
Schwinger et al. \cite{1} inferred the standard set of Maaxwell-Lorentz Equations (S-MLEs) by combining three ingredients are here noted in the opening paragraph of introduction to this paper. Here we wish to note further that if we proceed with the method we adopted for gravitational case in the previous section, we would obtain a set of New Maxwell-Lorentz Equations (N-MLEs) given in Table 2, which produce the same physical effects as the S-MLEs. Thus S-MLEs and N-MLEs represent are mere two different mathematical representations of a single physical theory: The Standard Electromagnetic Theory. The simplest way to get the results of Table-2 from Table-1 and vice-versa is the substitution 
\begin{subequations} \label{eq:7}
\begin{align}
\rho_g\,\leftrightarrow \,\rho_e,  \quad  \mathbf{g} \,\leftrightarrow \,\mathbf{E}, \quad  -4\pi G \,\leftrightarrow \,\frac{1}{\epsilon_0} \\
\mathbf{j}_g\,\leftrightarrow \,\mathbf{j}_e, \quad \mathbf{b} \,\leftrightarrow \,\mathbf{B} \quad  c_g \,\leftrightarrow \, c,  \quad  -\mu_{0g} \leftrightarrow \,  \mu_{0}   \\
m_g\,\leftrightarrow \,q,  \quad \phi_g  \,\leftrightarrow \, \phi_e, \quad  \mathbf{A}_g\,\leftrightarrow \,\mathbf{A}_e   \\ 
\text{g-MLEs of MG } \,\leftrightarrow \text{S-MLEs} \\  
\text{g-MLEs of HG } \,\leftrightarrow \text{N-MLEs}
\end{align}
\end{subequations}
 \begin{table}[h]
\begin{center}
   \begin{tabular}{ | l | l |}
    \hline
    Standard MLEs    & New MLEs  \\ \hline
    $\mathbf{\nabla}\cdot\mathbf{E} = \frac{\rho_e}{\epsilon_{0}}$ & $\mathbf{\nabla}\cdot\mathbf{E} = \frac{\rho_e}{\epsilon_{0}}$ \\ \hline
    $\mathbf{\nabla}\cdot\mathbf{B} = 0$ & $\mathbf{\nabla}\cdot\mathbf{B} = 0$ \\ \hline
    $\mathbf{\nabla}\times\mathbf{B}\,=\,+\,\mu_{0}\mathbf{j}_e\,+\,\frac{1}{c^2}\frac{\partial \mathbf{E}}{\partial t}$ & $\mathbf{\nabla}\times\mathbf{B}\,=\,-\,\mu_{0}\mathbf{j}_e\,-\,\frac{1}{c^2}\frac{\partial \mathbf{E}}{\partial t}$  \\ \hline 
    $\mathbf{\nabla}\times\mathbf{E}\,=\,-\,\frac{\partial \mathbf{B}}{\partial t}$ & $\mathbf{\nabla}\times\mathbf{E}\,=\,+\,\frac{\partial \mathbf{B}}{\partial t}$  \\ \hline
    $\frac{d\mathbf{p}}{dt}\,=\,q\left[\mathbf{E}\,+\,\mathbf{u}\times \mathbf{B}\right]$ & $\frac{d\mathbf{p}}{dt}\,=\,q\left[\mathbf{E}\,-\,\mathbf{u}\times \mathbf{B}\right]$  \\ \hline 
   $\mathbf{B}\,=\,+\mathbf{\nabla}\times \mathbf{A}_e$ & $\mathbf{B}\,=\,-\mathbf{\nabla}\times \mathbf{A}_e$ \\ \hline
     $\mathbf{E}\,=\,-\,\mathbf{\nabla}\phi_{e} \,-\,\frac{\partial \mathbf{A}_e}{\partial t}$ & $\mathbf{E}\,=\,-\,\mathbf{\nabla}\phi_{e} \,-\,\frac{\partial \mathbf{A}_e}{\partial t}$ \\ \hline
\end{tabular}
\end{center}
\caption{Two Physically Equivalent Sets of Maxell-Lorentz Equations (in SI units) with Different Mathematical look. The symbols have their usual meanings.}
    \label{tab:truthTables}   
\end{table}
It is interesting to note that, Behera and Barik \cite{12} have recently obtained the four sets of equations listed in Table-1 and Table-2, from the well established principle of local phase (or gauge) invariance of quantum field theory, which corroborate our present findings.
\section{Discussions}
The author wish to note that Einstein was unaware of Heaviside's extension of Newtonian gravity to time-dependent sources and fields in the spirit of Faraday-Maxwell theory of electromagnetism, otherwise he would have restrained himself to make the following remark on Newtonian gravity before the 1913 congress of natural scientists in Vienna \cite{43}, viz., 
\begin{quote}
After the un-tenability of the theory of action at distance had thus been proved in the domain of electrodynamics, confidence in the correctness of Newton's action-at-a-distance theory of gravitation was shaken. One had to believe that Newton's law of gravity could not embrace the phenomena of gravity in their entirety, any more than Coulomb's law of electrostatics embraced the theory of electromagnetic processes.
\end{quote} 
This is because of the place of publication of Heaviside's work and the poor state of communication system prevalent at that time. Had it been widely known, by some means, Heaviside's gravity would have played the same role, on equal footing as Maxwell's electromagnetic theory did, in the development of Einsteinian relativity as noted by Schwinger. It is also interesting to find Heaviside to have discussed the propagation of gravitational waves carrying energy momentum in terms of gravitational analogue of electromagnetic Heaviside-Poynting's theorem. McDonald \cite{3} who reported Heaviside's gravity in the form of Maxwellian gravity and described it as a low velocity, weak-field approximation to general relativity in response to the question, `Why $``c"$ for gravitational waves?' Recently Ummarino and Gallerati \cite{44} derived Heaviside's Gravity in the form of Maxwellian Gravity from Einstein's GR by certain linearization procedure in the weak field and slow motion approximations, which corroborates MacDonald's remark on Heaviside's gravity. In Ummarino and Gallerati's version of Heaviside-Maxwellian Gravity one obtains $c_g = c$. Unfortunately, there are several other versions of gravito-Maxwell-Lorentz equations in different linearized versions of GR, which have recently been critically examined in ref. \cite{4} in connection with the questions of the value $c_g$, the correspondence principle, the spin of graviton and shown that they are not unique and unambiguous in the context of GR. Einstein's assumption of the equality of gravitational mass with inertial mass has also been shown to be violated without sacrificing Galileo's experimental law of universality of free fall in 
refs. \cite{2,4}. Moreover the theoretical objections against the spin-1 theory of gravity to which HMG belongs has also been refuted by Behera and Barik recently \cite{12}, who obtained the equations of HMG using the principle of local phase invariance of quantum field theory. Since the results reported here agree with the theoretical considerations of several authors\cite{2,4,11,12,25,26,26,27,28,29,30,31,32,33,34,35,36}, Heaviside's gravitational theory seems to be a self consistent field theory of gravity in the spirit of Faraday-Maxwell theory of electromagnetism.   \\ 
As regards its agreement with experimental data, refs. \cite{37,38,39,40,41,42} may be consulted for the explanations of (a) the Mercury's perihelion advance, (b) bending of photon's path in the gravitational field of the Sun, and (c) Shapiro time delay in a gravitational field. Concerning the recent experimental detection of gravitational waves and their explanation within the framework of vector gravitational theory like HMG, the works of Mead \cite{45,46,47} and Hilborn \cite{42} are noteworthy. Thus it seems Rothman \cite{48} is right in stating, {\it ``One hesitates to make rash claims, but Heaviside's article may well have been the world's first serious scientific paper to treat gravitational waves"}. Unfortunately, Heaviside's theory of gravitomagnetism and gravitational waves is not widely circulated. Therefore it has not received as much attention as it merits. One finds rare or no mention of Heaviside's name in leading literature on gravitomagnetic effects and gravitational waves, although Heaviside's prediction of gravitomagnetic effects and gravitational waves in 1893, is almost 20 years ahead of Einstein's prediction of gravitational waves \cite{49,50}. 
In his final remark on Carstoiu's \cite{26,27} 1969 suggestions for gravitational waves that has already been considered by Heaviside in 1893, Brillouin \cite{28} rightly stated, ``{\it It is very strange that such an important paper had been practically ignored for so many years, but the reader may remember that Heaviside was the forgotten genius of physics, abandoned by everybody except a few faithful friends.}"  
\section{Conclusions}
Inspired by Einstein-Infeld's \cite{51} philosophy of scientific advancement, viz.,``{\it To raise new questions, new possibilities, to regard old problems from a new angle, requires creative imagination and marks real advance in science}”, and following Schwinger's Galilio-Newtonian reletivistic  formalism for inferring Maxwell-Lorentz equations of classical electromagnetism, we inferred two sets of gravito-Maxwell-Lorentz equations of classical gravitoelectromagnetism form three plausible assumptions: (1) Newton's law of gravitostatics, (2) the relativity principle of Galileo and Newtonian and (3) the existence of gravitational waves which travel in vacuum with some finite speed $c_g$. One of the two sets of gravito-Maxwell-Lorentz equations corresponds to Heaviside's Gravity (HG) of 1893 and the other set represents what we call Maxwellian Gravity (MG). Since HG and MG, produce identical physical effects, they  represent a single physical theory called Heaviside-Maxwellian Gravity (HMG). 
We also report a correction to the gravito-Lorentz force law speculated by Heaviside. Our results are in conflict with the existing scientific belief that gravitomagnetic effects and gravitational waves are absent in the domain of Galileo-Newtonian physics. The natural emergence HMG from other advanced and well established principles of physics, such as the Lorentz invariance of physical laws, the principle of gauge invariance of quantum field theory, is also noted, which established its theoretical consistency and fixed the value of $c_g$ uniquely at $c$, the speed of light in vacuum. The application of HMG in some form or other in reproducing certain experimentally verified general relativistic results has also been noted. The theory should, therefore, not be discarded, otherwise such rejection amounts to the rejection of the well established principles which are at the foundation of HMG. 
We hope our results may revive some interest in HMG to reexamine and explore it further for its utility or futility in the description of certain physical phenomena beyond Newton's gravitostatics to bring Heaviside's gravity, in either of the two forms presented here, to its logical, mathematical and physical conclusion in future. The Lorntz-invariant formulation of HMG reported elsewhere has room for further development to make generally co-variant. Physicists should never cease testing the well founded and  self consistent basic theories, out of curiosity that new physics could exist beyond the ``accepted" picture.


\begin{thebibliography}{}

\bibitem{1} J. Schwinger, L. L. DeRaad, Jr., K. A. Milton, Wu-yang Tsai, Classical Electrodynamics (Perseus Books, Reading, Massachusetts, 1998) pp. 8-12. 
\bibitem{2} H. Behera, P. C. Naik, Gravitomagnetic Moments and Dynamics of Dirac (spin 1/2) Fermions in Flat Spape-time Maxwellian Gravity, Int. J. Mod. Phys. A, {\bf 19}, 4207-4229 (2004). We named our relativistic gravity as Maxwellian Gravity since J. C. Maxwell [J.C Maxwell, Phil. Trans. Roy. Soc. London {\bf 155}, 459-512 (1865), sec. 82: Note on the Attraction of Gravitation] first tried to construct a field theory of gravity analogous to Classical Electromagnetic theory and did not work on the matter further as ``He was dissatisfied with his results because the potential energy of a static configuration is always negative but he felt this should be re-expressible as an integral over field energy density which, being the square of the gravitational field, is positive \cite{3}. 
\bibitem{3} K. T. McDonald, Answer to Question $\#49$. Why $c$ for gravitational waves? Am. J. Phys., {\bf 65} 591-592 (1997). 
\bibitem{4} H. Behera, Comments on gravitoelectromagnetism of Ummarino and Gallerati in “Superconductor in a weak static gravitational field” vs other versions. Eur. Phys. J. C., {\bf 77}, 822 (2017). \url{https://doi.org/10.1140/epjc/s10052-017-5386-4} 
\bibitem{5} O. Heaviside, A Gravitational and Electromagnetic Analogy, Part I, The Electrician, {\bf 31} 281-282 (1893); 
\bibitem{6} O. Heaviside,A Gravitational and Electromagnetic Analogy, Part II, The Electrician, {\bf 31} 359 (1893). 
\bibitem{7} O. Heaviside, \emph{Electromagnetic Theory}, vol.1 (The Electrician Printing and Publishing Co., London, 1894) p. 455-465. 
\bibitem{8} O. Heaviside, \emph{Electromagnetic Theory} (Dover, New York, 1950),  Appendix B, p. 115-118. (See also the quotation in the Introduction of this book.) 
\bibitem{9} O. Heaviside, \emph{Electromagnetic Theory}, vol. 1, 3rd Ed. (Chelsea Publishing Company, New York, N. Y., 1971) p.455-466.
\bibitem{10} An unedited copy of the original Heaviside's article, except that some formulas and all vector equations have been converted to modern notation, is reproduced in \cite{11} below, p. 189-202.
\bibitem{11} O. Jefimenko, \emph{Causality, electromagnetic induction, and gravitation : a different approach to the theory of electromagnetic and gravitational fields}, (Electret Scientific Company, Star City, West Virginia, 2000), 2nd Ed. In this book, Jefimenko has also obtained the equations of Maxwellian Gravity from the consideration of causality principle.
\bibitem{12} H. Behera and N.Barik, A New Set of Maxwell-Lorentz Equations and Rediscovery of Heaviside-Maxwellian (Vector) Gravity from Quantum Field Theory, (2018). Submitted to EPJ Plus (Under Review). \url{https://arxiv.org/abs/1810.04791}   
\bibitem{13} B. P. Abbott et al., GW170814: A Three-Detector Observation of Gravitational Waves from a Binary Black Hole Coalescence, Phys. Rev. Lett. {\bf 119}, 141101 (2017).
\bibitem{14} B. P. Abbott et al., GW170104: Observation of a 50-Solar-Mass Binary Black Hole Coalescence at Redshift 0.2, Phys. Rev. Lett. {\bf 118}, 221101 (2017). 
\bibitem{15} B. P. Abbott et al., GW151226: Observation of Gravitational Waves from a 22-Solar-Mass Binary Black Hole Coalescence, Phys. Rev. Lett. {\bf 116}, 241103 (2016). 
\bibitem{16} B. P. Abbott et al., Observation of Gravitational Waves from a Binary Black Hole Merger, Phys. Rev. Lett. {\bf 116}, 061102 (2016).
\bibitem{17} I. Ciufolini, E. Pavlis, et al., Test of General Relativity and Measurement of the Lense-Thirring Effect with Two Earth Satellites, Science, {\bf 279}, 2100-2103 (1998).
\bibitem{18} I. Ciufolini, E. C. Pavlis, A confirmation of the general
relativistic prediction of the Lense–Thirring effect, Nature, {\bf 431},  958-960 (2004).
\bibitem{19} I. Ciufolini, Dragging of inertial frames, Nature, {\bf 449}, 41-47 (2007).
\bibitem{20} L. Iorio, How accurate is the cancellation of the first even zonal harmonic of the geopotential in the present and future LAGEOS-based Lense-Thirring tests?, Gen. Rel. Grav. {\bf 43}, 1697-1706 (2011).  
\bibitem{21} L. Iorio, The impact of the orbital decay of the LAGEOS satellites on the frame-dragging tests, Adv. in Space Res., {\bf 57}, 493–498 (2016). 
\bibitem{22} C. W. F. Everitt, et al., Gravity Probe B: Final Results of a Space Experiment to Test General Relativity, Phys. Rev. Lett., {\bf 106}, 221101 (2011).
\bibitem{23} C. M. Will, Finally, results from Gravity Probe B, Physics {\bf 4}, 43 (2011).
\bibitem{24} C. W. F. Everitt, et al., The Gravity Probe B test of general relativity, Class. Quantum Grav. {\bf 32}, 224001 (2015).  
\bibitem{25} O. Jefimenko, \textit{Gravitation and Cogravitation: Developing Newton's Theory of Gravitation to its Physical and Mathematical Conclusion}, Electret Scientific Company, Star City (2006).   
\bibitem{26} J. Carstoiu, Les deux champs de gravitation et propagation des ondes gravifiques, Compt. Rend. {\bf 268}, 201-263 (1969). 
\bibitem{27} J. Carstoiu, Nouvelles remarques sur les deux champs de gravitation et propagation des ondes gravifiques, Compt. Rend. {\bf 268}, 261-264 (1969).
\bibitem{28} L. Brillouin, \emph{ Relativity Reexamined} (Academic Press, New York, 1970).
\bibitem{29} H. G. L. Cosster, J. R. Shepanski, Gravito-inertial fields and relativity, J . Phys. A (Gen. Phys.), {\bf 2}, Ser. 2, 22-27 (1969).
\bibitem{30} H. G. L. Cosster, J. R. Shepanski, Gravito-inertial fields and the theory of a neutral particle, J . Phys. A (Gen. Phys.), {\bf 2}, Ser. 2, 257-261 (1969).
\bibitem{31} D. D. Cattani, Linear equations for the gravitational field, Nuovo Cimento B, Serie 11 {\bf 60B} 67-80 (1980).
\bibitem{32} S. Demir, Space-time algebra for the generalization of gravitational field equations, Pramana - J Phys, {\bf 80},811 (2013). \url{https://doi.org/10.1007/s12043-013-0516-5} 
\bibitem{33} J. A. Heras, An axiomatic approach to Maxwell's equations, Eur. J. Phys. {\bf 37} 055204 (2016).  
\bibitem{34} G. G. Nyambuya, Fundamental Physical Basis for Maxwell-Heaviside Gravitomagnetism, Journal of Modern Physics, {\bf 6}, 1207-1219 (2015). 
\bibitem{35} D. H. Sattinger, Gravitation and Special Relativity, J. Dyn. Diff. Equat., {\bf 27} 1007-1025 (2015). 
\bibitem{36} R. S. Vieira, H. B. Brentan, Covariant theory of gravitation in the framework of special relativity, Eur. Phys. J. Plus, {\bf 133}, 165 (2018). \url{https://doi.org/10.1140/epjp/i2018-11988-9} 
\bibitem{37} R. J. Kennedy, Planetary motion in a Retarded Newtonian Field, Proc. N. A. S. {\bf 15}, 744 (1929) . \url{https://doi.org/10.1073/pnas.15.9.744} 
\bibitem{38} D. W. Sciama, On the Origin of Inertia, Monthly Notices of the Royal Astronomical Society, {\bf 113}, 34-42 (1953).  
\bibitem{39} A. Singh, Experimental Tests of the Linear Equations for the Gravitational Field, Lettere Al Nuovo Cimento, {\bf 34}, 193-196 (1982). 
\bibitem{40} W. D. Flanders and G. S. Japaridze, Photon deflection and precession of the periastron in terms of spatial gravitational fields, Class. Quantum Grav. {\bf 21}, 1825-1831 (2004).
\bibitem{41} V. N. Borodikhin, Vector Theory of Gravity, Gravitation and Cosmology, {\bf 17}, 161-165 (2011). 
\bibitem{42} R. C. Hilborn, Gravitational waves from orbiting binaries without general relativity, Am.J. Phys. {\bf 86}, 186 (2018). \url{https://doi.org/10.1119/1.5020984} 
\bibitem{43} R. Torreti, \emph{Relativity and Geometry},(Dover Pub. Inc., New York, 1996), p. 130.
\bibitem{44} G.A. Ummarino, A. Gallerati, Superconductor in a weak static gravitational field. Eur. Phys. J. C. {\bf 77}, 549 (2017). 
\bibitem{45} C. Mead, Gravitational Waves in G4v (2015). \url{https://authors.library.caltech.edu/59770/}
\bibitem{46} M. Isi, A.J. Weinstein, C. Mead, M. Pitkin, Detecting beyond-Einstein polarizations of continuous gravitational waves, Phys. Rev. D {\bf 91}, 082002 (2015). 
\bibitem{47} Watch the seminar talk of C. Mead in YouTube: \url{https://www.youtube.com/watch?v=XdiG6ZPib3c} 
\bibitem{48} T. Rothman, The Secret History of Gravitational Waves, American Scientist, {\bf 106}, 96 (2018). DOI: 10.1511/2018.106.2.96. 
\bibitem{49} A. Einstein, N\"{a}herungsweise Integration der Feldgleichungen der Gravitation, Sitzungber.
Preuss. Akad. Wiss. Berlin, part 1, 688 (1916).  
\bibitem{50} A. Einstein, \"{U}ber Gravitationswellen, Sitzungber. Preuss. Akad. Wiss. Berlin, part 1, 154
(1918).
\bibitem{51} A. Einstein, \& L.Infeld, \emph{The Evolution of Physics} (Cambridge Univ. Press, London, 1967), p. 95. 


\end{thebibliography}
\end{document}